\documentclass[12pt]{article}
\usepackage{times,graphicx}
\begin{document}

\noindent
{\large\em Oxford Research Encyclopedia of Physics}$\,$ {\large (2019)}

\noindent
\hrulefill

\noindent
{\large\bf Solar Wind Origin}

\noindent
Steven R.\  Cranmer

\noindent
{\small
Subject: Cosmology and Astrophysics
$\,$\dotfill$\,$
DOI: 10.1093/acrefore/9780190871994.013.18 \\
Post-Referee Author Draft}

\vspace*{-0.06in}
\noindent
\hrulefill

\vspace*{0.10in}
\noindent
{\large\bf Summary and Keywords}

The Sun continuously expels a fraction of its own mass in the form
of a steadily accelerating outflow of ionized gas called the
``solar wind.''
The solar wind is the extension of the Sun's hot (million-degree Kelvin)
outer atmosphere that is visible during solar eclipses as the bright
and wispy corona.
In 1958, Eugene Parker theorized that a hot corona could not exist for
very long without beginning to accelerate some of its gas into
interplanetary space.
After more than half a century, Parker's idea of a gas-pressure-driven
solar wind still is largely accepted, although many questions remain
unanswered.
Specifically, the physical processes that heat the corona have not yet
been identified conclusively, and the importance of additional wind
acceleration mechanisms continue to be investigated.
Variability in the solar wind also gives rise to a number of practical
``space weather'' effects on human life and technology, and there is
still a need for more accurate forecasting.
Fortunately, recent improvements in both observations (with telescopes
and via direct sampling by space probes) and theory (with the help of
ever more sophisticated computers) are leading to new generations of 
predictive and self-consistent simulations.
Attempts to model the origin of the solar wind are also leading to new
insights into long-standing mysteries about turbulent flows,
magnetic reconnection, and kinetic wave-particle resonances.


\vspace*{-0.06in}
\noindent
\hrulefill

\vspace*{0.10in}
\noindent
{\large\bf Introduction}

The solar wind is a tenuous gas surrounding the Sun that contains
neutral atoms, positively charged ions, and free electrons.
These particles accelerate away from the solar surface and fill the
majority of the volume of the solar system, thus contributing to
the fact that outer space is not a pure vacuum.
The ever-expanding solar wind carries with it some of the Sun's
complex and multipolar magnetic field, which becomes stretched out
along the mostly radial streamlines followed by the particles
(see, e.g., Figure~1).
Far from the Sun, the typical outflow speeds of solar wind particles
range between about 250 and 800 km/s.
The elemental composition of the solar wind is similar to that of
the Sun's interior: by mass, roughly 78\% hydrogen, 20\% helium,
and 2\% other elements---mostly oxygen, carbon, iron, neon,
magnesium, nitrogen, and silicon.
Collectively, these atoms, ions, and electrons compose a magnetized
fluid that is an outer extension of the Sun's atmosphere.

\newpage
As it accelerates away from the Sun, the solar wind carries information
about its origin in the solar corona.
This information is encoded in the statistical distribution of particle
speeds that varies continuously as a function of time and radial
distance away from the Sun.
The bulk outflow velocity of the solar wind is the mean value of this
distribution.
There is always a nonzero random spread about the mean (i.e., the
temperature) as well as asymmetries such as ``skewness'' that determine
how momentum and energy are transported through the gas.
Deep-space probes have measured these particle distributions
{\em in~situ} (i.e., locally, or in place) since the early 1960s,
and they also provide a direct sampling of the electric and magnetic
fields through which the particles travel.
In addition, telescopes of various types observe solar photons that are
scattered or emitted by the particles.
These photons carry useful diagnostic information about the particle
velocities and the magnetic field in the solar wind that is highly
complementary to the {\em in~situ} data.

\begin{figure}[!t]
\vspace*{0.00in}
\hspace*{-0.02in}
\includegraphics[width=6.44in]{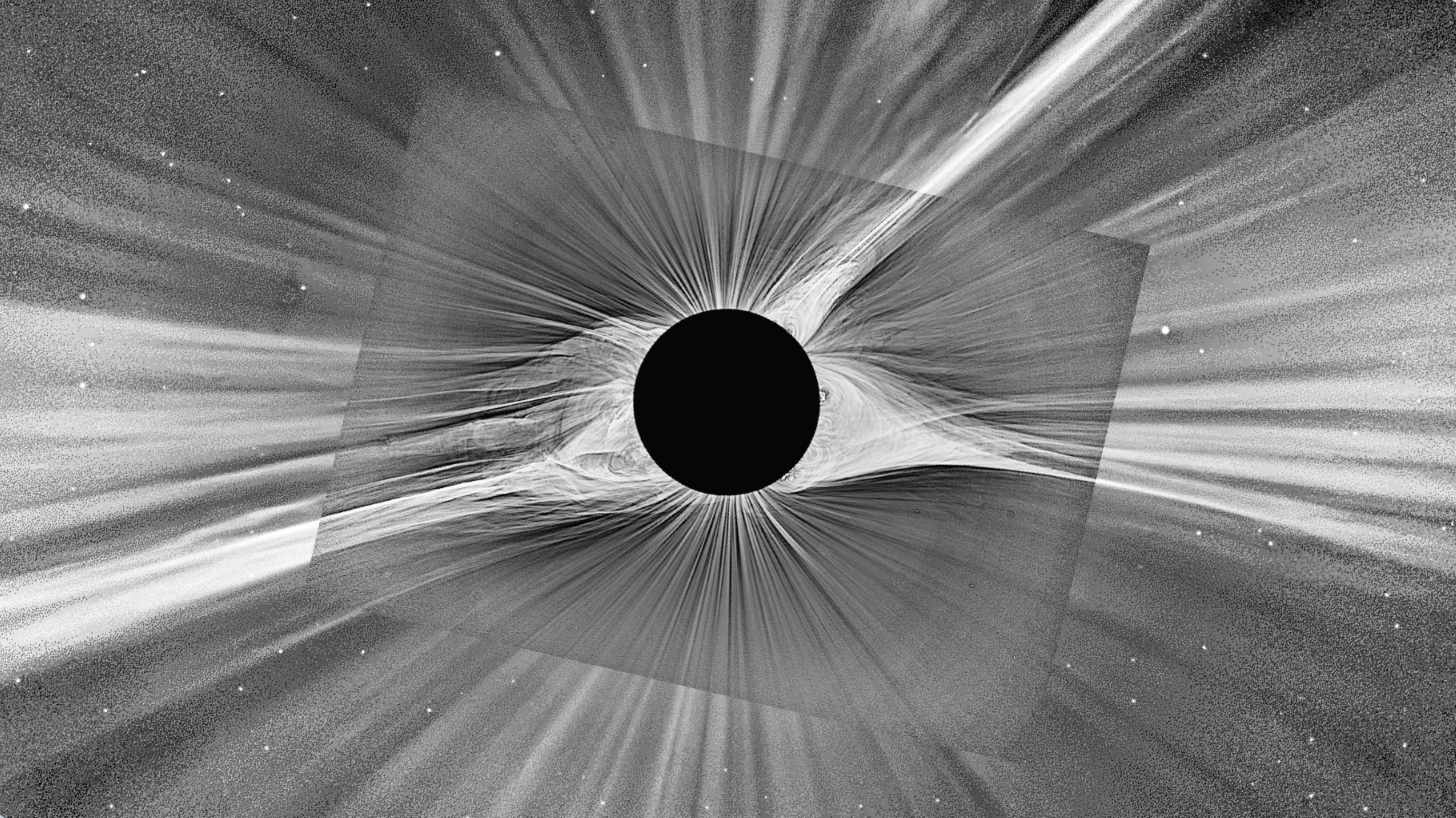}

\vspace*{0.07in}
\noindent
\baselineskip=12.0pt
{\small
{\bf Figure 1:}
Perhaps the best way to ``see'' the solar wind is to block out the
bright solar disk.
This allows the much fainter solar corona, consisting of scattered
light from charged particles in million-degree outer atmosphere
of the Sun, to become observable.
This is a composite, processed illustration of the structures visible
during the total solar eclipse of August 21, 2017 (adapted from original
images obtained and processed by M.\  Druckm\"{u}ller, P.\  Aniol,
Z.\  Druckm\"{u}llerov\'{a}, J.\  Hoderov\'{a}, P.\  \v{S}tarha, 
and S.\  Habbal).
Scattered light traces out lines of magnetic force, which in turn
have been stretched out by the expansion of the solar wind.
}

\vspace*{0.03in}
\end{figure}

The Sun loses approximately $10^{36}$ particles per second
(i.e., about a billion kilograms per second) to the solar wind, which
means that it is expected to lose roughly 0.02\% of its total mass
over its 10 billion year main-sequence lifetime.
This is a small fraction, but the solar wind is believed to have been
even more important to the Sun's evolution in other ways.
For example, the solar wind carries away some of the Sun's angular
momentum (see, e.g., Weber \& Davis, 1967).
Because of the magnetized nature of the fluid, the solar wind exerts
a magnetic torque that is thought to have slowed the Sun's rotation
from an initial period of about 1 day to its current rotation period
of about 26 days.
Also, when the Sun was very young and still surrounded by a dense
accretion disk, its rate of mass loss was thought to be several
orders of magnitude larger than it is now.
The correspondingly denser ``protosolar wind'' may have been responsible
for eroding away the primordial atmospheres of the inner planets
(Lammer et al., 2012).

Additional motivations for studying the solar wind are often invoked
from the standpoint of either pragmatism (i.e., space weather
forecasting) or as an accessible laboratory for fundamental physics
research.
On the pragmatic side, there is an ever-increasing need to understand
how the Sun affects both technology and society.
When the variable solar wind impacts the Earth's magnetosphere, it
can interrupt communications, disrupt power grids, damage satellites,
and threaten the safety of humans in space
(Feynman \& Gabriel, 2000; Koskinen et al., 2017).
On the other hand, the study of the physical processes at work in the
solar wind has established a baseline of knowledge that helps us
understand the inner workings of more distant astrophysical systems.
For example, theoretical insights about plasma heating mechanisms in
the solar wind have been applied to other stars (Cranmer \& Saar, 2011),
the interstellar medium (Spangler, 1991),
galaxy clusters (Parrish et al., 2012),
and supermassive black holes (Sironi \& Narayan, 2015).

\vspace*{0.06in}
\noindent
{\large\bf Observational Overview}

\vspace*{0.06in}
\noindent
{\bf Historical Discovery}

Since ancient times, it has been possible to observe {\em both} the
launching of the solar wind (i.e., the corona during a total eclipse)
and its eventual termination in the Earth's atmosphere (i.e.,
the shimmering aurora borealis) with the unaided eye.
From the late 19th century to the early 20th century, there arose
considerable speculation that these phenomena were associated with
one another.
Carrington (1859) and others noted that the large solar flare of
September 1, 1859 was soon followed by intense geomagnetic storms
(rapid fluctuations in the Earth's magnetic field) and disruptions
along telegraph lines.
Ennis (1878) speculated that the solar corona, the terrestrial aurora,
and tails of comets were all similar manifestations of the same kind of
``streaming forth of electricity.''
Birkeland (1908), after collecting data for many years on polar
expeditions, provided additional evidence that both auroral activity
and geomagnetic storms ``...should be regarded as manifestations of an
unknown cosmic agent of solar origin.''
Serviss (1909) compared the ``sheaves of light emanating from the poles''
of the Sun's corona to the field lines associated with the poles of
a bar magnet.
He also hypothesized that ``...there exists a direct solar influence
not only upon the magnetism, but upon the weather of the earth.''

From the time of these earliest thoughts about a correlation between
the Sun and the Earth's local space environment, it took several
decades to work out the precise means of causation.
Chapman (1918) and others proposed that the Sun emits sporadic
clouds or beams of charged particles into the vacuum of space.
Hulburt (1937) proposed that bursts of ultraviolet radiation from
the Sun were the primary means of exciting the aurora and geomagnetic
storms.
Pikel'ner (1950) realized that solar electrons can much more easily
escape the Sun's gravity than the heavier protons, and he speculated
about the buildup of a net charge in the solar atmosphere.
These ideas all had the benefit of being partially correct, but
they did not yet fully come to grips with the idea of a {\em plasma}
that is made up of both positively and negatively charged particles,
and which behaves like a volume-filling gas or fluid.
Biermann (1951) realized that the existence of cometary ion tails
required there be a continuous outflow of ``corpuscular radiation''
(i.e., plasma) throughout the solar system.

At the same time as the above solar--terrestrial connections were
being assembled, there was also evidence slowly building that the
Sun's corona is extremely hot.
During the total solar eclipse of August 7, 1869, Harkness and
Young first observed a bright green spectral line in the coronal
emission at a wavelength of 530.3 nm.
Because spectroscopists had not yet identified any chemical
element that emits at this wavelength, Lockyer (1869) reported
that these observations were
``...{\em{bizarre}} and puzzling to the last degree!''
There was speculation that these observations implied the existence
of a new element (``coronium''), and in subsequent eclipses
a few dozen other mysterious coronal lines were discovered.
The identification of these spectral features was a problem that
persisted for at least a half-century.
Eventually, Grotrian (1939) and Edl\'{e}n (1943) applied insights
from the newly developed quantum theory to determine that these lines
were coming from unusually high ionization states of iron, calcium,
and nickel.
Alfv\'{e}n (1941) assembled additional pieces of evidence to confirm
that the corona must be a highly ionized plasma with a temperature
of roughly $10^6$ K.

Parker (1958) brought together the idea of a hot corona with
Biermann's concept of a continuously outflowing collection of particles,
and he found them to be inescapably linked with one another.
Parker realized that the high coronal temperature produced a strong
outward force due to the gradient of gas pressure.
This force is sufficient to counteract gravity and drive a time-steady
accelerating flow of plasma that he named the {\em solar wind.}
The initial model of Parker (1958) made the simplifying assumption
of a constant coronal temperature, which is justifiable because
the high conductivity of a hot plasma leads to efficient spatial
diffusion (i.e., smearing out) of temperature fluctuations (Chapman, 1957).
Parker's insightful calculation led to an implicit transcendental
equation that contained both the radial outflow speed $v$ and the
radial distance $r$.
Finding a closed-form analytic solution to this kind of equation
(i.e., solving explicitly for $v$ as a function of $r$) had to wait
almost a half-century for the development of new mathematical functions
designed for these types of equations (see, e.g., Cranmer, 2004).

Parker's solution for the solar wind was ``transonic.''
In other words, it involved a slow and subsonic expansion near the
Sun, which begins to exceed the speed of sound at a critical radius
in the corona, and then keeps accelerating as a supersonic flow at
larger distances.
This was not the only possible solution of Parker's equations, and
Chamberlain (1961) proposed that one of the other solutions---a
{\em solar breeze} that always remains subsonic---was a more plausible
one for nature to choose.
There was also a controversy regarding Parker's assumption that the
plasma behaves as a continuous fluid.
Models such as Chamberlain's were based on a kinetic (i.e.,
collisionless particle-by-particle escape) approach that is still
used in modeling planetary exospheres.
At the time, these two approaches gave different answers, and it
was not clear which one was correct.

Prior to resolving these issues, Parker (1963) and others began to
extend the original model to account for a radially varying
coronal temperature.
Depending on how and where the corona is heated, the high conductivity
of the plasma allows for the temperature to increase to a maximum
value in the corona, then decrease with increasing distance.
Figure~2 illustrates this kind of model with an analytic temperature
that obeys the dictates of its high conductivity both in the low
corona (see, e.g., Rosner et al., 1978) and at large distances
(Chapman, 1957).
Hotter temperatures produce a critical radius closer to the Sun
(i.e., the gas-pressure gradient starts overcoming gravity sooner)
and a higher solar wind speed in interplanetary space.

\begin{figure}[!t]
\vspace*{0.00in}
\hspace*{0.95in}
\includegraphics[width=4.40in]{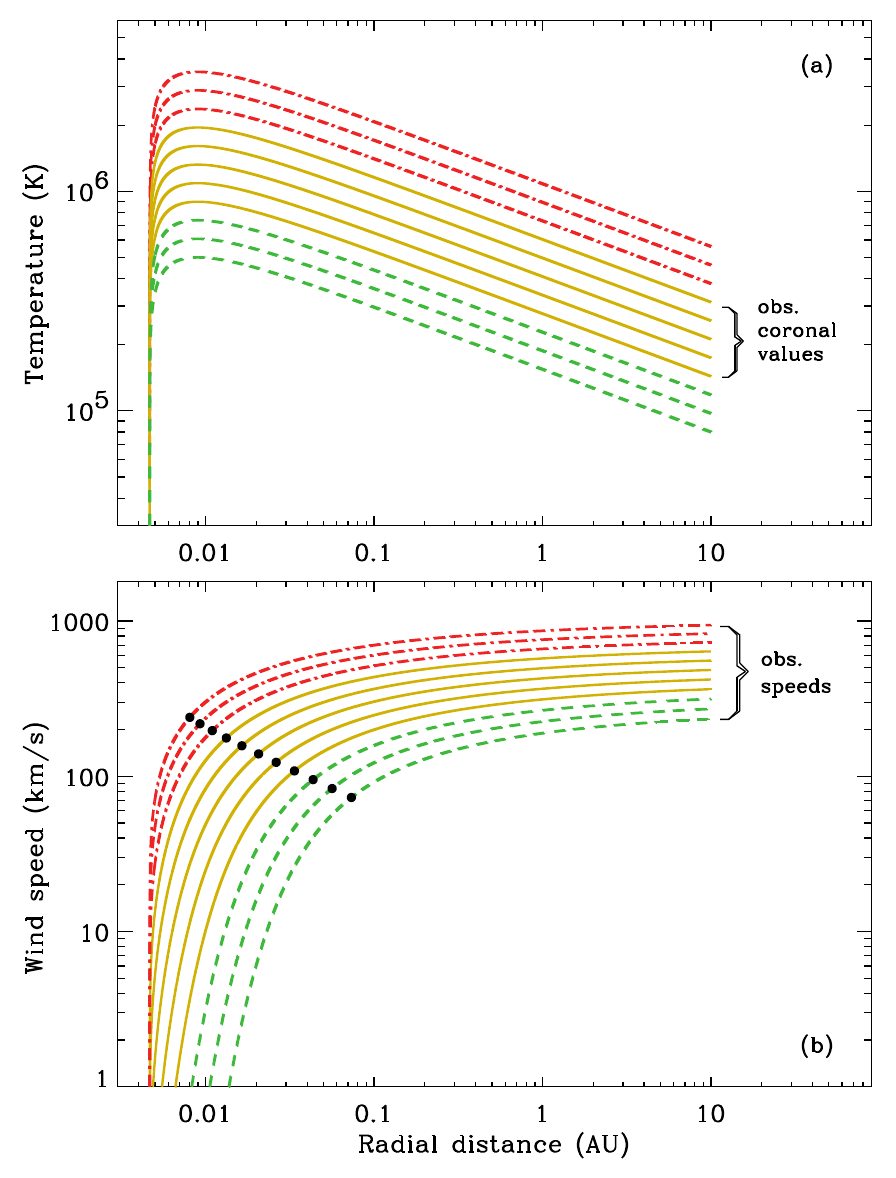}

\vspace*{-0.01in}
\noindent
\baselineskip=12.0pt
{\small
{\bf Figure 2:}
Example solutions to Parker's solar wind equations:
(a) a series of empirically constrained estimates for the 
temperature of the solar corona, and
(b) the resulting radial dependence of outflow speed corresponding
to each temperature curve.
Parker's critical radii are labeled with black circles.
Note that, in the original theory, the full range of 
observed solar wind speeds (i.e., red, gold, and green curves)
comes about only because of a relatively large range of coronal
temperatures.
However, the observed range of temperatures only spans the middle
(gold) range of curves.
Thus, additional physics is needed to explain the fastest (red)
and slowest (green) solar wind streams.
}

\vspace*{0.03in}
\end{figure}

\vspace*{0.06in}
\noindent
{\bf The Space Age: Verification and Exploration}

It was fortuitous that Parker's first theoretical model of the
solar wind was published at the dawn of the Space Age,
because the community only had a few years to wait for deep-space
missions that ventured outside the Earth's magnetosphere.
The first {\em in~situ} detection of solar wind particles came
from a series of Russian {\em Lunik} and {\em Venera} probes between
1959 and 1961 (Gringauz et al., 1962).
Also in 1961, the American satellite {\em Explorer 10} measured
plasma velocities and densities just outside the magnetopause.
However, all of these measurements were relatively brief.
The ultimate confirmation of the existence of Parker's solar wind---i.e.,
the fact that it is always present and is always flowing at supersonic
speeds---was provided by {\em Mariner 2,}
which was sent to Venus in 1962 (Neugebauer \& Snyder, 1966).
Several months worth of continuous data revealed the presence of
alternating dense, low-speed (250--500 km/s)
streams and tenuous, high-speed (500--900 km/s) streams.

Soon after its discovery, additional large-scale structure in the
interplanetary medium was detected.
Wilcox \& Ness (1965) analyzed data from {\em IMP~1} (i.e.,
{\em Explorer 18,} which was renamed as the first {\em Interplanetary
Monitoring Platform}) to reveal that the sign of the radial component
of the magnetic field reverses its polarity every few days.
This pattern of oppositely directed ``magnetic sectors'' also tends to
recur roughly every 27 days, which is close to the Sun's equatorial
rotation period.
Figure~3a illustrates the current understanding of these sectors;
they each map back to discrete source-regions in the solar corona of
the same polarity.
The high conductivity of plasma in the corona and solar wind ensures
that the particles and the lines of magnetic force remain closely
tied to one another, so that the entire system rotates together.
Alfv\'{e}n (1957) and Parker (1958) realized that this effect would lead
to radially flowing streams of particles winding up the magnetic
field lines into rotating Archimedean spirals.
This is analogous to the way that a spinning lawn-sprinkler emits
radial jets of water that are observed to be twisted into spiral-shaped
streaklines.
Slower wind streams are more tightly wound than faster streams.
This means that when fast and slow streams are emitted at neighboring
longitudes, they eventually interact with one another to form
{\em cororating interaction regions} (CIRs).
Figure~3b shows that high-density regions (i.e., compressions) occur
when fast streams catch up with slow streams, and that low-density
regions (rarefactions) occur when the slow streams lag behind.

\begin{figure}[!p]
\vspace*{-0.25in}
\hspace*{1.20in}
\includegraphics[width=4.10in]{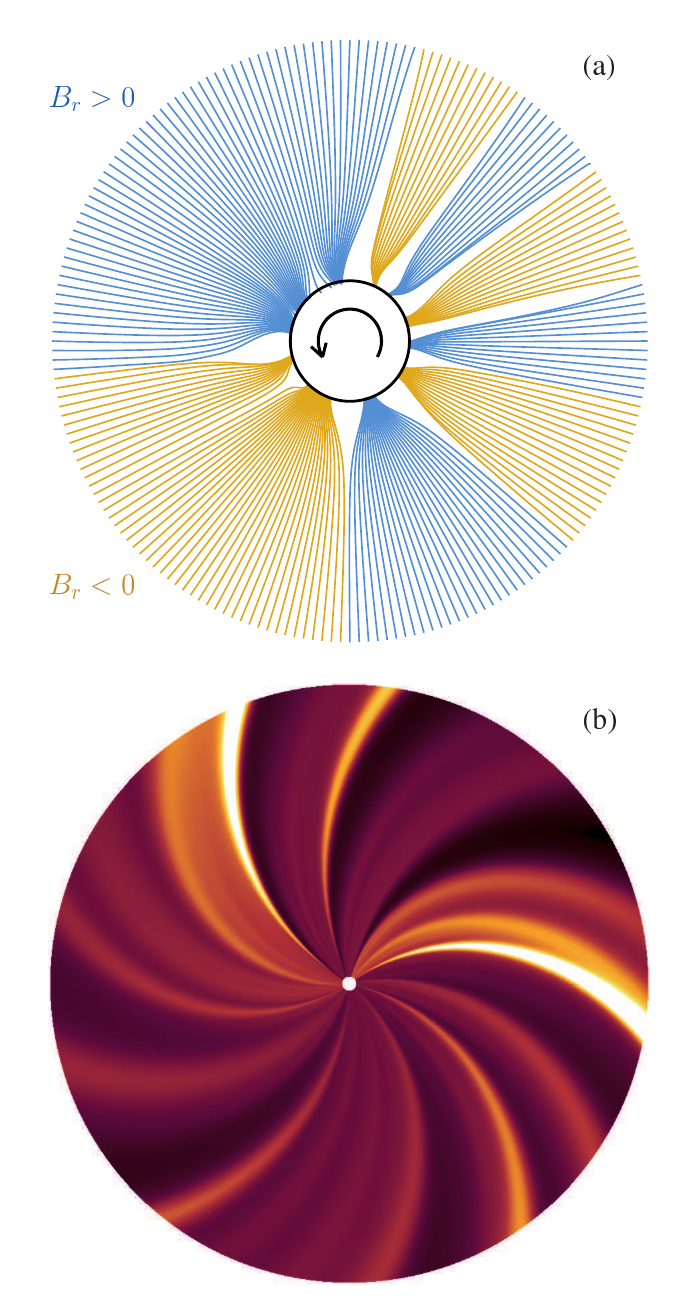}

\vspace*{-0.03in}
\noindent
\baselineskip=12.0pt
{\small
{\bf Figure 3:}
(a) Looking down from the Sun's north polar axis, magnetic field lines
that fill the equatorial plane can be traced down to smaller source
regions on the solar surface.
Field lines of positive (blue) and negative (gold) polarity were
computed for the time period June--July 2007 by the
Magnetohydrodynamics Around a Sphere (MAS) code (Linker et al., 1999).
(b) As the wind accelerates out in the equatorial plane,
fast and slow streams interact to form CIRs.
Dark [light] colors illustrate low [high] density structures that
follow the so-called ``Parker spiral'' field.
The MAS model from panel (a) was used as input to a kinematic
simulation that evolves the plasma from 0.02 to 1 AU (see, e.g.,
Riley \& Lionello, 2011).
}

\vspace*{0.03in}
\end{figure}

Spatial and temporal variability in the solar wind extends down to
very small scales.
Belcher \& Davis (1971) discovered that so-called Alfv\'{e}n waves
(i.e., transverse oscillations in the magnetic field that do not
disturb the density or pressure like sound waves do) are ubiquitous
in the solar wind.
Nonlinear interactions between different types of fluctuations
create a chaotic ``turbulent cascade'' in which large eddies are
broken up into ever-smaller eddies (see, e.g., Bruno \& Carbone, 2013).
In fact, the eddies keep cascading over at least 8 orders of magnitude
in size until they find themselves at the scales on which {\em individual}
charged particles interact with one another and with the magnetic field.
In regions of low density---where inter-particle collisions become
infrequent---these effects manifest themselves as departures from
thermal equilibrium (i.e., individual particle species drifting past
one another at different speeds, and temperatures measured along the
magnetic field being unequal to temperatures measured transverse to
the field; see, e.g., Marsch, 2006).

In the decades since the solar wind's discovery, there have been many
other deep-space missions that have studied it in more detail.
The earliest missions tended to stay close to the Earth's orbit
at a heliocentric distance of 1 astronomical unit (AU).
The twin {\em Helios} probes plunged inside the orbit of Mercury
to 0.28 AU (Marsch, 1991).  The {\em Voyager} probes are now past
120 AU---i.e., approximately three times more distant than
Pluto---and are still sending back data (Richardson et al., 2017).
In the 1990s, {\em Ulysses} became the first spacecraft to pivot
away from the solar system's ecliptic plane and measure the
solar wind over the north and south poles of the Sun (Marsden, 2001).
The {\em Ulysses, SOHO} ({\em{Solar and Heliospheric Observatory}}),
{\em Wind,} and {\em ACE} ({\em{Advanced Composition Explorer}})
missions contained instruments that were able to measure the precise
abundances and ratios of ionization charge states for multiple
elements heavier than hydrogen and helium (e.g., Zurbuchen, 2007).
Because these ``composition'' signatures do not evolve as rapidly as
other properties of the plasma (e.g., temperature, total density, and
flow speed), they are useful tracers of the coronal origins of
solar wind streams.

A survey of solar wind measurements would not be complete without
also including telescopic remote-sensing.
With Lyot's development of the coronagraph in the 1930s, it became
possible to observe scattered light from the solar corona at times
other than total solar eclipses (see, e.g., Billings, 1966).
Coronagraphic measurements in broadband visible light are sensitive
to the number free electrons along various lines-of-sight through
the corona.
Thus, it has become possible to use techniques derived from medical
tomography to determine the three-dimensional distribution of 
electron density (Frazin et al., 2010).
Rapid-cadence sequences of coronagraphic images have also revealed
the presence of low-level density fluctuations that drift to larger
radii as a function of time.
These structures are believed to act like passive ``leaves in the wind''
and thus probe the acceleration profile of the flow
(Sheeley et al., 1997; Cho et al., 2018).
Lastly, there are also coronagraphic instruments that disperse
the incoming light into a spectrum.
This allows hundreds of bright coronal emission lines (including
those that were initially mistaken for ``coronium'') to be used as
diagnostics of elemental abundances, plasma velocity distributions,
and turbulent fluctuations.
In the regions of the corona undergoing the fewest inter-particle
collisions, the observations show similar departures from thermal
equilibrium as are seen by the {\em in~situ} instruments
(see, e.g., Kohl et al., 2006).

\vspace*{0.06in}
\noindent
{\large\bf Theoretical Understanding}

\vspace*{0.06in}
\noindent
{\bf What Heats the Solar Corona?}

The source of solar wind acceleration proposed by
Parker (1958, 1963)---i.e., an outwardly directed gas-pressure gradient
force due to the existence of the hot corona---is still believed
to be responsible for much of what we observe.
Early questions about (1) the viability of the transonic solution to the
conservation equations, and (2) mutual consistency between fluid-based
and kinetic approaches to modeling the flow, were resolved more or less
in Parker's favor (Velli, 1994; Echim et al., 2011).
Thus, the most substantial remaining problem is how to
produce the required million-degree coronal temperatures.

The ultimate source of energy for heating the solar corona is the
convection zone that churns just below the surface.
Only about 1\% of the kinetic energy in the rising/falling motions
of convective ``granules'' needs to be delivered up to the corona
and subsequently converted into heat.
Because more highly magnetized regions of the Sun appear to receive
more coronal heating, it is widely believed that the upward
delivery of energy must involve temporary storage in the form of
magnetic energy.
Thus, a reasonable way of expressing the maximum available energy is
to evaluate the upward component of the {\em Poynting flux} (i.e.,
the energy transfer per unit area, per unit time, by a magnetic
field in motion).
An upward Poynting flux at the solar surface can be achieved by
either (1) transverse jostling of mostly-vertical field lines, or
(2) upward flow of mostly-horizontal field lines.
Of course, for nearly all regions on the actual Sun, the field is
so twisted and tangled---and the underlying motions are so complex---that
the Poynting flux probably has nontrivial contributions from both effects.

The irreversible conversion of magnetic energy to heat is most efficient
when there arise structures on very small spatial scales.
In other words, an increase in the random thermal motions of particles
is most likely to occur when there is activity on scales commensurate
with the (microscopic) random inter-particle collisions.
For the corona, there is still no agreement on how this happens;
see recent reviews by Parnell \& De Moortel (2012),
Schmelz \& Winebarger (2015), and Cranmer et al.\  (2017).
There are two overall schools of thought that depend on the relative
values of two key {\em time-scales.}
First, there is the so-called Alfv\'{e}n travel-time $t_{\rm A}$,
which is the time that it takes a small wavelike fluctuation to
traverse a significant distance along the coronal magnetic field.
Second, there is the photospheric footpoint time-scale $t_{\rm ph}$,
which is a characteristic time over which the convective motions
can make major changes to the field at the bottom of the corona.
Thus, there are two limiting cases:
\begin{enumerate}
\item[1.]
If $t_{\rm ph} > t_{\rm A}$, then the corona has ample time to
respond to the slow underlying motions.
The coronal magnetic field relaxes into a succession of increasingly
braided and stressed configurations.
At any arbitrary time in such a system, there will be a selection of
locations at which the magnetic complexity exceeds a stability
threshold.
Each of these locations will then undergo a rapid, explosive burst
of current dissipation known as magnetic reconnection
(Parker 1972, 1988).
These bursts of heating are often called ``nanoflares,''
and the models that describe these kinds of quasi-static equilibria
are called {\bf direct-current (DC)} heating theories.
These mechanisms are invoked more often for closed magnetic loops than
for the open field lines that feed the solar wind.
\item[2.]
On the other hand, if $t_{\rm ph} < t_{\rm A}$, then the rapid footpoint
driving creates fluctuations that propagate along the coronal magnetic
field lines in the form of waves, shocks, and turbulent eddies
(see, e.g., Osterbrock 1961; Roberts 2000).
Small spatial scales can be generated when the waves interact with
background gradients in density (e.g., reflection or refraction at
sharp boundaries) or when neighboring waves travel at different
speeds through an inhomogeneous plasma (e.g., ``phase mixing'').
Eventually, the waves damp out, either via particle-particle collision
effects (i.e., viscosity, thermal conductivity, or electrical
resistivity) or via wave-particle interactions that occur when there
are departures from thermal equilibrium.
The damped wave energy is converted to heat, and the models that
utilize this source of energy are often called
{\bf alternating-current (AC)} heating theories.
\end{enumerate}

\noindent
One source of ongoing difficulty with the above ideas is that the
observed time-scales often take on different values in different regions.
Statistically, both $t_{\rm ph}$ and $t_{\rm A}$ tend to have
broad distributions that overlap with one another, so it is unclear
whether there are very many strictly AC or DC regions.
Also, some observational clues once believed to be unique to one
paradigm seem to have the characteristics of the opposite paradigm.
For example, high-resolution ultraviolet imaging has seen signs of
both the bursty nanoflare events and tightly braided magnetic fields
characteristic of DC theories.
However, these tend to coincide with rapid and fluctuating flows
that are expected from AC theories (see, e.g.,
Cirtain et al., 2013; Testa et al., 2013).
This kind of coronal activity can be described using the unifying
language of a turbulent cascade (see, e.g., Milano et al., 1997;
van Ballegooijen et al., 2014; Velli et al., 2015).

\newpage
\noindent
{\bf What Else Accelerates the Wind?}

Figure 2 shows that there are high-speed wind streams (i.e.,
exceeding 600 km/s) that are difficult to explain with Parker's
original gas-pressure driving theory alone.
These fast streams tend to contain relatively high fluxes of
Alfv\'{e}n waves that propagate away from the Sun.
Thus, Belcher (1971) and Alazraki \& Couturier (1971) proposed
that these oscillations may exert an additional outward-directed
{\em wave pressure} that provides extra acceleration to the wind.
Several years earlier, Bretherton \& Garrett (1968) showed that
waves propagating through an inhomogeneous medium can exert
a nondissipative net force on the fluid.
This force can be expressed as the gradient of a pressure-like
quantity, and the effect is somewhat analogous to the way that
electromagnetic waves carry momentum and exert pressure on matter.
Recent models of the solar wind that include wave pressure have
been successful in predicting the properties of high-speed solar
wind streams (see, e.g., Cranmer et al., 2007; Ofman, 2010;
Oran et al., 2013).

In the lowest-density parts of the corona---in which inter-particle
collisions are rare---departures from thermal equilibrium can
provide additional ways of accelerating the solar wind.
There can be ``two-temperature'' distributions of particle velocities,
in which the spread of random velocities perpendicular to the
magnetic field lines is larger than the spread parallel to
the field lines (Marsch, 2006).
In this case, particles that flow away from the Sun undergo a
similar {\em magnetic mirror} effect as particles in the Earth's
Van Allen radiation belts.
As individual particles travel from strong to weak magnetic-field
regions, their perpendicular gyrations become redirected to be
parallel to the (mostly radially oriented) field lines.
This provides extra kinetic energy for the outflowing solar wind.

The slowest streams of the solar wind are seen to flow away from
the Sun at speeds down to about 250 km/s.
The original Parker (1958) theory would have demanded coronal
temperatures of only about 500,000 K to create a wind this slow,
but such low temperatures do not tend to be observed.
It is possible that these regions are the result of energy loss
in the stream-stream collisions that form CIRs.
Alternately, they may be associated with highly distorted bundles
of magnetic field lines; i.e., far from the radially oriented
rays assumed in Parker's spherically symmetric model.
These geometrical effects are described in more detail in the
section titled {\em Forecasting the Wind Speed.}
Lastly, Kasper et al.\  (2007) discovered an intriguing fact about
the elemental composition of the slowest solar wind:
the relative abundance of helium decreases rapidly as one
approaches the low-speed limit of 250 km/s from above.
No solar wind at all is observed with speeds below the value at
which the helium abundance would be extrapolated to zero.
Thus, there is speculation that the presence of helium is a crucial
piece of the solar wind acceleration puzzle.

\vspace*{0.06in}
\noindent
{\bf What Determines the Sun's Mass Loss?}

Parker's (1958) theory, and its many extensions, is largely successful
in predicting the acceleration and speed of the solar wind.
However, it does so without putting any clear limits on how many
particles are ejected per unit time.
Observations and models have shown that the full Sun emits
roughly $2 \times 10^{-14}$ solar masses per year in the solar
wind at the minimum of the sunspot activity cycle, and roughly
$3 \times 10^{-14}$ solar masses per year at the maximum of the cycle
(Wang, 1998).
Note that the Sun's interior mass is also depleted at a slightly higher
rate (i.e., $6.7 \times 10^{-14}$ solar masses per year) by the
conversion of hydrogen to helium via nuclear fusion.
However, this latter rate of mass depletion is observable only in the
form of the solar luminosity (i.e., mass--energy release as photons)
and not as a component of the solar wind.

Given that both the solar magnetic field and the total rate of
coronal heating vary by orders of magnitude over the 11-year activity
cycle, it is somewhat surprising that the mass-loss rate varies by
only about 50\%.
However, the energy deposited as coronal heating is very efficiently
transported away from where it originates, both inwards (in the form
of thermal conduction) and outwards (as the solar wind).
These effects essentially smear out the effects of any large changes
in the heating. 
In other words, this acts as a kind of ``thermostat'' that limits
the variability of the coronal temperature and pressure
(see, e.g., Rosner et al., 1978; Hammer, 1982; Leer et al., 1982).
The thermostat effect is a consequence of energy conservation, and
the requirement for a steady-state balance between gains
(heating) and losses (both transport and actual cooling via the
emission of photons) is what sets plasma density---and thus also
the mass-loss rate---to vary as weakly as it does.

Although the above idea of thermal energy balance has been the
dominant explanation for the Sun's mass loss, another idea has been
suggested.
High-resolution observations of the solar chromosphere (a relatively
cool layer between the photosphere and corona) have revealed the
presence of a wide range of narrow, short-lived features known
variously as spicules, jets, fibrils, and mottles.
Many of these structures show a rapid upward surge of plasma,
followed by much (or all) of that plasma falling back down.
It has been suggested that some of this plasma continues traveling
up and is eventually heated to become the corona and the base of the
solar wind (see, e.g., Moore et al., 2011; McIntosh, 2012).
This is related to the idea of {\em interchange reconnection} in the
low corona, in which magnetic loops emerge from below the solar
surface, undergo magnetic reconnection with neighboring regions,
and drive jet-like surges of plasma up into the corona and solar
wind (Fisk et al., 1999; Yang et al., 2013).
Although coronal jets are indeed observed, they appear as bright and
narrow features in images because they only occupy a small fraction
of the coronal volume at any time.
Thus, it remains uncertain whether they can be responsible for the
majority of the plasma mass of the corona and solar wind.

\vspace*{0.06in}
\noindent
{\bf Forecasting the Wind Speed}

As summarized in the {\em Introduction,} there is a definite
societal need to be able to make predictions of the future state
of the solar wind.
Putting aside violent events such as solar flares and coronal
mass ejections (CMEs), it is also important to know how the Sun
produces its ``ambient'' distribution of fast and slow streams.
In turn, this requires knowing which magnetic structures in the
corona connect to which kinds of wind streams at 1~AU.
The highest-speed streams appear to be connected to the largest
{\em coronal holes} on the Sun's surface (Cranmer, 2009).
These are the darkest and least active regions of the Sun, but they
expand out majestically over the north and south poles over much of
the solar cycle (see Figure~1).
On the other hand, the slow solar wind is associated with a range
of observed magnetic structures---e.g., cusp-like streamers often
seen at low latitudes, fan-like active regions associated with
sunspots, quiet regions with only intermittent open fields, and
the outer edges of coronal holes (see, e.g., Abbo et al., 2016).
Because these structures all undergo substantial time variability,
their precise relative contributions to the slow wind are still
not known.

No matter the dominant magnetic sites of origin for solar wind streams,
it is now well-known that there is a relationship between the speed
of the wind and the {\em flux-tube geometry} of the field-lines
that connect that stream down to the solar surface.
Levine et al.\  (1977) and Wang \& Sheeley (1990) independently
discovered that the wind speed at 1~AU is correlated inversely
with the amount of transverse flux-tube expansion between the
surface and a reference point in the mid-corona
(see also Arge \& Pizzo, 2000).
Bundles of field lines that trumpet out abruptly from a small
source region on the Sun generally produce slow wind.
Bundles of field lines that undergo only moderate lateral expansion
(such as the central regions of coronal holes) generally produce
fast wind.
The physical origin of this effect is suspected to be related to
the emission of Alfv\'{e}n waves at the base of the corona
(Kovalenko 1981; Wang \& Sheeley, 1991).
If every region on the surface produces the same energy flux
(i.e., power per unit area), then regions that expand greatly---from
small patches on the surface to large areas in the outer corona---will
receive lower total amounts of wave energy than the regions that
map down to larger patches on the surface and do not expand as much.
Thus, the low-expansion regions with the highest amounts of coronal
wave energy would then be heated more vigorously (from, e.g., the
turbulent cascade of those waves), and thus be subject to higher
degrees of solar wind acceleration.
The relationship between flux-tube expansion and solar wind speed
forms the backbone of much practical space-weather forecasting
(see, e.g., Riley et al., 2015).

\vspace*{0.06in}
\noindent
{\large\bf Conclusions and Future Prospects}

Although there have been substantial accomplishments in both
observations and theory over the past few decades, there is much more
to be done in order to reach a complete understanding of the origin
of the solar wind.
Advances in computational efficiency have allowed the community to
build increasingly sophisticated numerical simulations
of the corona and solar wind (e.g.,
Gressl et al., 2014; Carlsson et al., 2016; Chen et al., 2017;
Yalim et al., 2017; Gombosi et al., 2018; Shoda et al., 2018),
and the use of these kinds of models in real-time forecasting is
on the horizon.
Multidimensional simulations are increasingly making use of
{\em data assimilation} (i.e., constant updating based on the most
recent observations) to improve their effectiveness and their
fidelity with nature.

In addition to theoretical advances, there are also new vistas in
observational capability coming soon.
The {\em Parker Solar Probe} (Fox et al., 2016) is expected to
revolutionize our conception of the inner heliosphere by performing
{\em in~situ} sampling closer to the Sun than any other prior space
mission.
The {\em Daniel K.\  Inouye Solar Telescope}
(DKIST; Tritschler et al., 2016) will be the largest-diameter
solar telescope in history, and it will measure the solar corona's
magnetic field at unprecedented sensitivity.
There are also plans in development for placing solar wind monitors
at various locations in the solar system---e.g., at the Earth-Sun L5
point or over the solar poles---that would help fill in huge
observational gaps and thus lead to more accurate space-weather
forecasting.

\noindent
\hrulefill

\vspace*{0.06in}
\parindent=0.00in
{\large\bf Further Reading}

Aschwanden, M. J. (2006).
{\em Physics of the solar corona: An introduction with
problems and solutions} (2nd ed.).
Chichester, UK: Springer--Praxis.

Dessler, A. J. (1967),
Solar wind and interplanetary magnetic field.
{\em Reviews of Geophysics, 5,} 1--41.

Golub, L., \& Pasachoff, J. M. (2010).
{\em The solar corona} (2nd ed.).
Cambridge: Cambridge University Press.

Hundhausen, A. J. (1972).
{\em Coronal expansion and solar wind.}
Berlin: Springer--Verlag.

Meyer-Vernet, N. (2007).
{\em Basics of the solar wind.}
Cambridge: Cambridge University Press.

Neugebauer, M. (1997),
Pioneers of space physics: A career in the solar wind.
{\em Journal of Geophysical Research, 102,} 26887--26894.

Parker, E. N. (2014),
Reminiscing my sixty year pursuit of the physics of the Sun
and the Galaxy.
{\em Research in Astronomy and Astrophysics, 14,} 1--14.

\vspace*{0.06in}
{\large\bf References}

Abbo, L., Ofman, L., Antiochos, S. K., Hansteen, V. H., Harra, L.,
Ko, Y.-K., ...~, Wang, Y.-M. (2016),
Slow solar wind: Observations and modeling.
{\em Space Science Reviews, 201,} 55--108.

Alazraki, G., \& Couturier, P. (1971),
Solar wind acceleration caused by the gradient of Alfv\'{e}n
wave pressure.
{\em Astronomy \& Astrophysics, 13,} 380--389.

Alfv\'{e}n, H. (1941),
On the solar corona.
{\em Arkiv f\"{o}r Matematik, Astronomi och Fysik (Band 27A), 25,} 1--23.

Alfv\'{e}n, H. (1957),
On the theory of comet tails.
{\em Tellus, 9,} 92--96.

Arge, C. N., \& Pizzo, V. J. (2000),
Improvement in the prediction of solar wind conditions using
near-real time solar magnetic field updates.
{\em Journal of Geophysical Research, 105,} 10465--10480.

Belcher, J. W. (1971),
Alfv\'{e}nic wave pressures and the solar wind.
{\em The Astrophysical Journal, 168,} 509--524.

Belcher, J. W., \& Davis, L., Jr. (1971),
Large-amplitude Alfv\'{e}n waves in the interplanetary medium, 2.
{\em Journal of Geophysical Research, 76,} 3534--3563.

Biermann, L. (1951),
Kometenschweife und solare Korpuskularstrahlung.
{\em Zeitschrift f\"{u}r Astrophysik, 29,} 274--286.

Billings, D. E. (1966).
{\em A guide to the solar corona.}
New York: Academic Press.

Birkeland, K. (1908).
{\em The Norwegian Aurora Polaris Expedition, 1902-–1903.}
New York and Christiania (now Oslo): H.\  Aschehoug \& Co.

Bretherton, F. P., \& Garrett, C. J. R. (1968),
Wavetrains in inhomogeneous moving media.
{\em Proceedings of the Royal Society A, 302,} 529--554.

Bruno, R., \& Carbone, V. (2013).
The solar wind as a turbulence laboratory.
{\em Living Reviews in Solar Physics, 10,} 2.

Carlsson, M., Hansteen, V. H., Gudiksen, B. V., Leenaarts, J.,
\& De Pontieu, B. (2016),
A publicly available simulation of an enhanced network region of
the Sun.
{\em Astronomy \& Astrophysics, 585,} A4.

Carrington, R. C. (1859),
Description of a singular appearance seen in the Sun on
September 1, 1859.
{\em Monthly Notices of the Royal Astronomical Society, 20,} 13--15.

Chamberlain, J. W. (1961),
Interplanetary gas, III: A hydrodynamic model of the corona.
{\em The Astrophysical Journal, 133,} 675--687.

Chapman, S. (1918),
An outline of a theory of magnetic storms.
{\em Proceedings of the Royal Society A, 95,} 61--83.

Chapman, S. (1957),
Notes on the solar corona and the terrestrial ionosphere.
{\em Smithsonian Contributions to Astrophysics, 2,} 1--12.

Chen, F., Rempel, M., \& Fan, Y. (2017),
Emergence of magnetic flux generated by a solar convective dynamo, I:
The formation of sunspots and active regions, and the origin of
their asymmetries.
{\em The Astrophysical Journal, 846,} 149.

Cho, I.-H., Moon, Y.-J., Nakariakov, V. M., Bong, S.-C., Lee, J.-Y.,
Song, D., ...~, Cho, K.-S. (2018),
2D solar wind speeds from 6 to 26 solar radii in solar cycle 24
by using Fourier filtering.
{\em Physical Review Letters,} in press, https://arXiv.org/abs/1806.08540

Cirtain, J. W., Golub, L., Winebarger, A. R., de Pontieu, B.,
Kobayashi, K., Moore, R. L., ...~, DeForest, C. E. (2013),
Energy release in the solar corona from spatially resolved magnetic
braids. {\em Nature, 493,} 501--503.

Cranmer, S. R. (2004),
New views of the solar wind with the Lambert W function.
{\em American Journal of Physics, 72,} 1397--1403.

Cranmer, S. R. (2009),
Coronal holes.
{\em Living Reviews in Solar Physics, 6,} 3.

Cranmer, S. R., Gibson, S. E., \& Riley, P. (2017),
Origins of the ambient solar wind: Implications for space weather.
{\em Space Science Reviews, 212,} 1345--1384. 

Cranmer, S. R., \& Saar, S. H. (2011),
Testing a predictive theoretical model for the mass loss rates of
cool stars.
{\em The Astrophysical Journal, 741,} 54.

Cranmer, S. R., van Ballegooijen, A. A., \& Edgar, R. J. (2007),
Self-consistent coronal heating and solar wind acceleration
from anisotropic magnetohydrodynamic turbulence.
{\em The Astrophysical Journal Supplement, 171,} 520--551.

Echim, M. M., Lemaire, J., Lie-Svendsen, {\O}. (2011),
A review on solar wind modeling: Kinetic and fluid aspects.
{\em Surveys in Geophysics, 32,} 1--70.

Edl\'{e}n, B. (1943),
Die deutung der emissionslinien im spektrum der sonnenkorona:
Mit 6 abbildungen.
{\em Zeitschrift f\"{u}r Astrophysik, 22,} 30--64.

Ennis, J. (1878),
The electric constitution of the solar system.
{\em Proceedings of the Academy of Natural Sciences of Philadelphia,}
January--April 1878, 102--118.

Feynman, J., \&  Gabriel, S. B. (2000),
On space weather consequences and predictions.
{\em Journal of Geophysical Research, 105,} 10543--10564.

Fisk, L. A., Schwadron, N. A., \& Zurbuchen, T. H. (1999),
Acceleration of the fast solar wind by the emergence of new magnetic flux.
{\em Journal of Geophysical Research, 104,} 19765--19772.

Fox, N. J., Velli, M. C., Bale, S. D., Decker, R., Driesman, A.,
Howard, R. A., ...~, Szabo, A. (2016),
The Solar Probe Plus mission: Humanity's first visit to our star.
{\em Space Science Reviews, 204,} 7--48.

Frazin, R. A., Lamy, P., Llebaria, A., \& V\'{a}squez, A. M. (2010),
Three-dimensional electron density from tomographic analysis of
LASCO--C2 images of the K-corona total brightness.
{\em Solar Physics, 265,} 19--30.

Gombosi, T. I., van der Holst, B., Manchester, W. B., \&
Sokolov, I. V. (2018),
Extended MHD modeling of the steady solar corona and the solar wind.
{\em Living Reviews in Solar Physics,} in press,
https://arXiv.org/abs/1807.00417

Gressl, C., Veronig, A. M., Temmer, M., Odstr\v{c}il, D.,
Linker, J. A., Miki\'{c}, Z., \& Riley, P. (2014),
Comparative MHD study of MHD modeling of the background solar wind.
{\em Solar Physics, 289,} 1783--1801.

Gringauz, K. I., Bezrukikh, V. V., Ozerov, V. D., \& Rybchinskii, R. E.
(1962), The study of interplanetary ionized gas, high-energy electrons
and corpuscular radiation of the sun, employing three-electrode charged
particle traps on the second Soviet space rocket.
{\em Planetary and Space Science, 9,} 103--107.

Grotrian, W. (1939),
Zur frage der deutung der linien im spektrum der sonnenkorona.
{\em Die Naturwissenschaften, 27,} 214.

Hammer, R. (1982),
Energy balance of stellar coronae, I: Methods and examples.
{\em The Astrophysical Journal, 259,} 767--778.

Hulburt, E. O. (1937),
Terrestrial magnetic variations and aurorae.
{\em Reviews of Modern Physics, 9,} 44--68.

Kasper, J. C., Stevens, M. L., Lazarus, A. J., Steinberg, J. T.,
\& Ogilvie, K. W. (2007),
Solar wind helium abundance as a function of speed and heliographic
latitude: Variation through a solar cycle.
{\em The Astrophysical Journal, 660,} 901--910.

Kohl, J. L., Noci, G., Cranmer, S. R., \& Raymond, J. C. (2006),
Ultraviolet spectroscopy of the extended solar corona.
{\em Astronomy \& Astrophysics Review, 13,} 31--157.

Koskinen, H. E. J., Baker, D. N., Balogh, A., Gombosi, T., Veronig, A.,
\& von Steiger, R. (2017),
Achievements and challenges in the science of space weather.
{\em Space Science Reviews, 212,} 1137--1157. 

Kovalenko, V. A. (1981),
Energy balance of the corona and the origin of quasi-stationary
high-speed solar wind streams.
{\em Solar Physics, 73,} 383--403.

Lammer, H., G\"{u}del, M., Kulikov, Y., Ribas, I., Zaqarashvili, T. V.,
Khodachenko, M. L., ...~, Fridlund, M. (2012),
Variability of solar/stellar activity and magnetic field and its
influence on planetary atmosphere evolution.
{\em Earth, Planets, and Space, 64,} 179--199.

Leer, E., Holzer, T. E., \& Fl{\aa}, T. (1982),
Acceleration of the solar wind.
{\em Space Science Reviews, 33,} 161--200.

Levine, R. H., Altschuler, M. D., \& Harvey, J. W. (1977),
Solar sources of the interplanetary magnetic field and solar wind.
{\em Journal of Geophysical Research, 82,} 1061--1065.

Linker, J. A., Miki\'{c}, Z., Biesecker, D. A., Forsyth, R. J.,
Gibson, S. E., Lazarus, A. J., ...~, Thompson, B. J. (1999),
Magnetohydrodynamic modeling of the solar corona during Whole Sun Month.
{\em Journal of Geophysical Research, 104,} 9809--9830.

Lockyer, J. N. (1869),
The recent total eclipse of the Sun.
{\em Nature, 1,} 14--15.

Marsch, E. (1991).
Kinetic physics of the solar wind plasma.
In R.\  Schwenn \& E.\  Marsch (Eds.),
{\em Physics of the inner heliosphere II: Particles, waves and
turbulence} (pp.\  45--133). Berlin: Springer-Verlag.

Marsch, E. (2006).
Kinetic physics of the solar corona and solar wind.
{\em Living Reviews in Solar Physics, 3,} 1.

Marsden, R. G. (2001),
The heliosphere after Ulysses,
{\em Astrophysics and Space Science, 277,} 337--347.

McIntosh, S. W. (2012),
Recent observations of plasma and Alfv\'{e}nic wave energy injection
at the base of the fast solar wind.
{\em Space Science Reviews, 172,} 69--87.

Milano, L. J., G\'{o}mez, D. O., \& Martens, P. C. H. (1997),
Solar coronal heating: AC versus DC.
{\em The Astrophysical Journal, 490,} 442--451.

Moore, R. L., Sterling, A. C., Cirtain, J. W., \& Falconer, D. A. (2011),
Solar X-ray jets, type-II spicules, granule-size emerging bipoles,
and the genesis of the heliosphere.
{\em The Astrophysical Journal, 731,} L18.

Neugebauer, M., \& Snyder, C. W. (1966),
Mariner 2 observations of the solar wind, 1: Average properties.
{\em Journal of Geophysical Research, 71,} 4469--4484.

Ofman, L. (2010),
Wave modeling of the solar wind.
{\em Living Reviews in Solar Physics, 7,} 4.

Oran, R., van der Holst, B., Landi, E., Jin, M., Sokolov, I. V.,
\& Gombosi, T. I. (2013),
A global wave-driven magnetohydrodynamic solar model with a unified
treatment of open and closed magnetic field topologies.
{\em The Astrophysical Journal, 778,} 176.

Osterbrock, D. E. (1961),
The heating of the solar chromosphere, plages, and corona by
magnetohydrodynamic waves.
{\em The Astrophysical Journal, 134,} 347--388.

Parker, E. N. (1958),
Dynamics of the interplanetary gas and magnetic fields.
{\em The Astrophysical Journal, 128,} 664--676.

Parker, E. N. (1963).
{\em Interplanetary dynamical processes.}
New York: Interscience Publishers.

Parker, E. N. (1972),
Topological dissipation and the small-scale fields in turbulent gases.
{\em The Astrophysical Journal, 174,} 499-510.

Parker, E. N. (1988),
Nanoflares and the solar X-ray corona.
{\em The Astrophysical Journal, 330,} 474--479.

Parnell, C. E., \& De Moortel, I. (2012),
A contemporary view of coronal heating.
{\em Philosophical Transactions of the Royal Society A, 370,} 3217--3240.

Parrish, I. J., McCourt, M., Quataert, E., \& Sharma, P. (2012),
The effects of anisotropic viscosity on turbulence and heat transport
in the intracluster medium.
{\em Monthly Notices of the Royal Astronomical Society, 422,} 704--718.

Pikel'ner, S. P. (1950),
On the theory of the solar corona.
{\em Bulletin of the Crimean Astrophysical Observatory, 5,} 34--58.

Richardson, J. D., Wang, C., Liu, Y. D., \v{S}afr\'{a}nkov\'{a}, J.,
N\v{e}me\v{c}ek, Z., \& Kurth, W. S. (2017),
Pressure pulses at Voyager 2: Drivers of interstellar transients?
{\em The Astrophysical Journal, 834,} 190.

Riley, P., Linker, J. A., \& Arge, C. N. (2015),
On the role played by magnetic expansion factor in the prediction of
solar wind speed.
{\em Space Weather, 13,} 154--169.

Riley, P., \& Lionello, R. (2011),
Mapping solar wind streams from the Sun to 1 AU: A comparison
of techniques.
{\em Solar Physics, 270,} 575--592.

Roberts, B. (2000),
Waves and oscillations in the corona: Invited review.
{\em Solar Physics, 193,} 139--152.

Rosner, R., Tucker, W. H., \& Vaiana, G. S. (1978),
Dynamics of the quiescent solar corona.
{\em The Astrophysical Journal, 220,} 643--665.

Schmelz, J. T., \& Winebarger, A. R. (2015),
What can observations tell us about coronal heating?
{\em Philosophical Transactions of the Royal Society A, 373,} 20140257.

Serviss, G. P. (1909).
{\em Curiosities of the sky: A popular presentation of the great
riddles and mysteries of astronomy.}
New York: Harper \& Brothers.

Sheeley, N. R., Wang, Y.-M., Hawley, S. H., Brueckner, G. E.,
Dere, K. P., Howard, R. A., ...~, Biesecker, D. A. (1997),
Measurements of flow speeds in the corona between 2 and 30 $R_{\odot}$.
{\em The Astrophysical Journal, 484,} 472--478.

Shoda, M., Yokoyama, T., \& Suzuki, T. K. (2018),
A self-consistent model of the coronal heating and solar wind acceleration
including compressible and incompressible heating processes.
{\em The Astrophysical Journal, 853,} 190.

Sironi, L., \& Narayan, R. (2015),
Electron heating by the ion cyclotron instability in collisionless
accretion flows, 1: Compression-driven instabilities and the
electron heating mechanism.
{\em The Astrophysical Journal, 800,} 88.

Spangler, S. R. (1991),
The dissipation of magnetohydrodynamic turbulence responsible for
interstellar scintillation and the heating of the interstellar
medium.
{\em The Astrophysical Journal, 376,} 540--555.

Testa, P., De Pontieu, B., Mart\'{\i}nez-Sykora, J., DeLuca, E.,
Hansteen, V., Cirtain, J., ...~, Weber, M. (2013),
Observing coronal nanoflares in active region moss.
{\em The Astrophysical Journal, 770,} L1.

Tritschler, A., Rimmele, T. R., Berukoff, S., Casini, R., Kuhn, J. R.,
Lin, H., ...~, the DKIST Team. (2016),
Daniel K.\  Inouye Solar Telescope: High-resolution observing of
the dynamic Sun.
{\em Astronomische Nachrichten, 337,} 1064--1069.

van Ballegooijen, A. A., Asgari-Targhi, M., \& Berger, M. A. (2014),
On the relationship between photospheric footpoint motions and
coronal heating in solar active regions.
{\em The Astrophysical Journal, 787,} 87.

Velli, M. (1994),
From supersonic winds to accretion: Comments on the stability of
stellar winds and related flows.
{\em The Astrophysical Journal, 432,} L55--L58.

Velli, M., Pucci, F., Rappazzo, F., \& Tenerani, A. (2015),
Models of coronal heating, turbulence, and fast reconnection.
{\em Philosophical Transactions of the Royal Society A, 373,}
20140262.

Wang, Y.-M. (1998).
Cyclic magnetic variations of the Sun.
In R.\  Donahue \& J.\  Bookbinder (Eds.),
{\em Tenth Cambridge Workshop on Cool Stars, Stellar Systems,
and the Sun,} ASP Conf.\  Ser.\ 154 (pp.\  131--151).
San Francisco: Astronomical Society of the Pacific.

Wang, Y.-M., \& Sheeley, N. R., Jr. (1990),
Solar wind speed and coronal flux-tube expansion.
{\em The Astrophysical Journal, 355,} 726--732.

Wang, Y.-M., \& Sheeley, N. R., Jr. (1991),
Why fast solar wind originates from slowly expanding coronal
flux tubes.
{\em The Astrophysical Journal, 372,} L45--L48.

Weber, E. J., \& Davis, L., Jr. (1967),
The angular momentum of the solar wind.
{\em The Astrophysical Journal, 148,} 217--227.

Wilcox, J. M., \& Ness, N. F. (1965),
Quasi-stationary corotating structure in the interplanetary medium.
{\em Journal of Geophysical Research, 70,} 5793--5805.

Yalim, M. S., Pogorelov, N., \& Liu, Y. (2017),
A data-driven MHD model of the global solar corona within Multi-Scale
Fluid-Kinetic Simulation Suite (MS-FLUKSS),
{\em Journal of Physics: Conference Series, 837,} 012015.

Yang, L., He, J., Peter, H., Tu, C.-Y., Chen, W.,
Zhang, L., ...~, Yan, L. (2013),
Injection of plasma into the nascent solar wind via reconnection
driven by supergranular advection.
{\em The Astrophysical Journal, 770,} 6.

Zurbuchen, T. H. (2007),
A new view of the coupling of the Sun and heliosphere.
{\em Annual Review of Astronomy \& Astrophysics, 45,} 297--338.

\end{document}